\documentclass[a4paper,11pt]{article}
\usepackage{amssymb}
\usepackage{amsfonts}
\usepackage{graphicx}
\usepackage{epstopdf}
\usepackage{dcolumn}
\usepackage{amsmath}
\usepackage{hyperref}
\usepackage{latexsym,bm}
\usepackage{geometry}

\def \be {\begin{equation}}
\def \ee {\end{equation}}
\def \bea {\begin{eqnarray}}
\def \eea {\end{eqnarray}}
\def \nn {\nonumber}

\def \a {\alpha}
\def \b {\beta}

\def \d {\delta}

\def \m {\mu}
\def \n {\nu}
\def \k {\kappa}

\def \s {\sigma}
\def \r {\rho}
\def \o {\omega}

\def \th {\theta}
\def \Th {\Theta}

\def \t {\tau}
\def \dag {\dagger}
\def \p {\partial}

\def\bd{\begin{document}}
\def\ed{\end{document}}
\def\nn{\nonumber}
\def\bea{\begin{eqnarray}}
\def\eea{\end{eqnarray}}
\let\bm=\bibitem
\let\la=\label

\def\N{{\cal N}}
\def\sst{\scriptscriptstyle}
\def\thetabar{\bar\theta}
\def\Tr{{\rm Tr}}
\def\one{\mbox{1 \kern-.59em {\rm l}}}

\def\a{\alpha}      \def\da{{\dot\alpha}}
\def\b{\beta}       \def\db{{\dot\beta}}
\def\c{\gamma}  \def\C{\Gamma}  \def\cdt{\dot\gamma}
\def\d{\delta}  \def\D{\Delta}  \def\ddt{\dot\delta}
\def\e{\epsilon}        \def\vare{\varepsilon}
\def\f{\phi}    \def\F{\Phi}    \def\vvf{\f}
\def\h{\eta}
\def\k{\kappa}
\def\l{\lambda} \def\L{\Lambda}
\def\m{\mu} \def\n{\nu}
\def\o{\omega}
\def\P{\Pi}
\def\r{\rho}
\def\s{\sigma}  \def\S{\Sigma}
\def\t{\tau}
\def\th{\theta} \def\Th{\Theta} \def\vth{\vartheta}
\def\X{\Xeta}
\def\z{\zeta}
\def\w{\wedge}
\def\u{\underline}
\def\hs{\hspace}

\def\cA{{\cal A}} \def\cB{{\cal B}} \def\cC{{\cal C}}
\def\cD{{\cal D}} \def\cE{{\cal E}} \def\cF{{\cal F}}
\def\cG{{\cal G}} \def\cH{{\cal H}} \def\cI{{\cal I}}
\def\cJ{{\cal J}} \def\cK{{\cal K}} \def\cL{{\cal L}}
\def\cM{{\cal M}} \def\cN{{\cal N}} \def\cO{{\cal O}}
\def\cP{{\cal P}} \def\cQ{{\cal Q}} \def\cR{{\cal R}}
\def\cS{{\cal S}} \def\cT{{\cal T}} \def\cU{{\cal U}}
\def\cV{{\cal V}} \def\cW{{\cal W}} \def\cX{{\cal X}}
\def\cY{{\cal Y}} \def\cZ{{\cal Z}}

\def\ua{\underline{\alpha}} \def\ubb{\underline{\beta}}
\def\ug{\underline{\gamma}}
\def\ub{\underline{\phantom{\alpha}}\!\!\!\beta}
\def\uc{\underline{\phantom{\alpha}}\!\!\!\gamma}
\def\um{\underline{\mu}} \def\un{\underline{\nu}}
\def\ud{\underline\delta}
\def\ue{\underline\epsilon}
\def\una{\underline a}\def\unA{\underline A}
\def\unb{\underline b}\def\unB{\underline B}
\def\unc{\underline c}\def\unC{\underline C}
\def\und{\underline d}\def\unD{\underline D}
\def\une{\underline e}\def\unE{\underline E}
\def\unf{\underline{\phantom{e}}\!\!\!\! f}\def\unF{\underline F}
\def\unm{\underline m}\def\unM{\underline M}
\def\unn{\underline n}\def\unN{\underline N}
\def\unp{\underline{\phantom{a}}\!\!\! p}\def\unP{\underline P}
\def\unq{\underline{\phantom{a}}\!\!\! q}
\def\unQ{\underline{\phantom{A}}\!\!\!\! Q}
\def\unH{\underline{H}}
\def\ul{\underline}

\def\As {{A \hspace{-6.4pt} \slash}\;}
\def\bs {{b \hspace{-6.4pt} \slash}\;}
\def\Ds {{D \hspace{-6.4pt} \slash}\;}
\def\ds {{\del \hspace{-6.4pt} \slash}\;}
\def\ss {{\s \hspace{-6.4pt} \slash}\;}
\def\ks {{ k \hspace{-6.4pt} \slash}\;}
\def\ps {{p \hspace{-6.4pt} \slash}\;}
\def\pas {{{p_1} \hspace{-6.4pt} \slash}\;}
\def\pbs {{{p_2} \hspace{-6.4pt} \slash}\;}

\def\Fh{\hat{F}}
\def\Vh{\hat{V}}
\def\Xh{\hat{X}}
\def\ah{\hat{a}}
\def\xh{\hat{x}}
\def\yh{\hat{y}}
\def\ph{\hat{p}}
\def\xih{\hat{\xi}}

\def\psit{\tilde{\psi}}
\def\Psit{\tilde{\Psi}}
\def\tht{\tilde{\th}}

\def\At{\tilde{A}}
\def\Qt{\tilde{Q}}
\def\Rt{\tilde{R}}
\def\Nt{\tilde{N}}

\def\at{\tilde{a}}
\def\st{\tilde{s}}
\def\ft{\tilde{f}}
\def\pt{\tilde{p}}
\def\qt{\tilde{q}}
\def\vt{\tilde{v}}
\def\nt{\tilde{n}}

\def\delb{\bar{\partial}}
\def\bz{\bar{z}}
\def\bD{\bar{D}}
\def\bB{\bar{B}}

\def\bk{{\bf k}}
\def\bl{{\bf l}}
\def\bp{{\bf p}}
\def\bq{{\bf q}}
\def\br{{\bf r}}
\def\bx{{\bf x}}
\def\by{{\bf y}}
\def\bR{{\bf R}}
\def\bV{{\bf V}}

\def\d{\delta}\def\D{\Delta}\def\ddt{\dot\delta}

\def\p{\partial} \def\del{\partial}
\def\xx{\times}
\def\uno{\mbox{1 \kern-.59em {\rm l}}}

\def\trp{^{\top}}
\def\inv{^{-1}}
\def\dag{{^{\dagger}}}

\def\pr{\prime}
\def\rar{\rightarrow}
\def\lar{\leftarrow}
\def\lrar{\leftrightarrow}

\title{Strong Subadditivity and Emergent Surface }
\author{
Bin Chen$^{1,2,3}$\footnote{bchen01@pku.edu.cn} and
Jiang Long$^{1}$\footnote{lj301@pku.edu.cn}
}
\date{}
\begin{document}
\maketitle
\begin{center}
{{\it
$^{1}$Department of Physics and State Key Laboratory of Nuclear Physics and Technology, Peking University, No. 5 Yiheyuan Rd, Beijing 100871, P.R.\! China\\
\vspace{2mm}
$^{2}$Collaborative Innovation Center of Quantum Matter, No. 5 Yiheyuan Rd, \\Beijing 100871, P.~R.~China\\
$^{3}$Center for High Energy Physics, Peking University, No. 5 Yiheyuan Rd, \\Beijing 100871, P.~R.~China
}
}
\vspace{10mm}
\end{center}

\date{}
\begin{abstract}

In this paper, we introduce two bounds which we call the Upper Differential Entropy and the Lower Differential Entropy for an infinite family of intervals(strips) in quantum field theory. The two bounds are equal provided that the theory is translational invariant and the entanglement entropy varies smoothly with respect to the interval. When the theory has a holographic dual, strong subadditivity of entanglement entropy indicates that there is always an emergent surface whose gravitational entropy is exactly given by the  bound.
 \end{abstract}
 \newpage

\section{Introduction}

Since its discovery, the Einstein's general relativity(GR) is the core to understand the connection between spacetime and geometry. The area of a surface, a geometric quantity, plays an quite important role to promote our understanding of the fundamental laws of physics. Even though its geometric meaning is lucid, its physical interpretation is intriguing. In short, there are two remarkable results on the area $\mathcal{A}$ of a surface in GR. The first one is the Bekenstein-Harking formula of a black hole entropy \cite{Bekenstein, Harking}
 \be
 S_{BH}=\frac{\mathcal{A}}{4G_N}\label{BH}
 \ee
    It relates the area of a Killing horizon to the thermal entropy of the system, and plays the key role in the black hole thermodynamics. How to understand the area law of the black hole entropy is one of most important questions in quantum gravity. It inspired people to propose  the holographic principle in quantum gravity.  The other one is the Ryu-Takayanagi(RT) formula for the holographic entanglement entropy \cite{Ryu0603}
 \be
 S_{EE}(I)=\mathop{\mbox{ext}}_{m\sim I}\frac{\mathcal{A}(m)}{4G_N}.\label{RT}
 \ee
It gives a simple prescription relating the entanglement entropy of a submanifold $I$ in a conformal field theory(CFT) to the area of an extremal bulk surface $m$ which is homologous to the boundary region $I$.

 The surfaces appear in (\ref{BH}) and (\ref{RT}) are either a Killing horizon or an extremal surface, so they are quite special in some sense. In the most general case, we may choose a time slice $\Sigma_{0}$ of a spacetime and a region $V\subseteq \Sigma_0$. Now the region $V$ and its complement $\bar{V}$ compose the time slice $\Sigma_{0}$. Their common boundary is denoted as $\partial V$. On one hand, we can define  a natural geometric quantity $\mathcal{A}(\partial V)$,  the area of $\partial V$. On the other hand, we cannot always relate it to a physical quantity, except in the above two cases\footnote{Some interesting efforts can be found in \cite{Bianchi, Balasubramanian}.}. This is an interesting phenomenon, as it indicates that our understanding of $\mathcal{A}(\partial V)$ is incomplete. The lesson from (\ref{BH}) and (\ref{RT}) tells us that there may be some generalized gravitational entropy\cite{Lewkowycz:2013nqa,Balasubramanian:2013rqa}
 \be
 S_{gr}=\frac{\mathcal{A}(\partial V)}{4G_N}\label{gr}
 \ee
 for arbitrary $\partial V$\footnote{To make the picture clear, we always illustrate our examples in the Einstein-Hilbert theory, but  we can easily generalize the arguments to other theories.}. Though the meaning of the formula (\ref{gr}) is not as clear as (\ref{BH}) and (\ref{RT}), there is interesting progress recently. In \cite{Balasubramanian1310}, the authors considered the gravitational entropy (\ref{gr}) of an arbitrary bulk curves in $AdS_3$, and related it to the so called ``Differential Entropy''
 \be
 E=\lim_{n\to \infty}\sum_{k=1}^n[S_{EE}(I_k)-S_{EE}(I_k\cap I_{k+1})]\label{RE}
 \ee
 in the boundary conformal field theory. Here the intervals $I_k(k=1,2,\cdots,n)$ are constructed as follows. We denote the $AdS_3$ coordinates as $t,x,z$, among which $t,x $ are also the boundary $CFT$ coordinates, $z$ represents  the extra dimension. A closed bulk curve $z=z(x)$ is assumed to be smooth. We divide the curve according to $n$ points $(x_k,z(x_k))(k=1,2,\cdots,n)$\footnote{Here we omit the coordinate $t$ as it is the same for the points in the bulk curve.}. When $k>n$, we impose the periodic condition: $k\sim k+n$. For each point labeled by $k$, we search for a boundary interval $I_k$ such that
 \begin{enumerate}
 \item The corresponding bulk extremal curve goes through the point $(x_k,z(x_k))$.
 \item The corresponding bulk extremal curve is tangent to the original curve $z=z(x)$ at $(x_k,z(x_k))$.
 \end{enumerate}
After finding the intervals $I_k$, the authors  calculated the differential entropy (\ref{RE}) and found a remarkable equality
\be
S_{gr}=E.\label{equality}
\ee
In \cite{Balasubramanian1310}, it was argued that the differential entropy is a measure of uncertainty about the state of a system left by an infinite family of local, finite-time observables. It seems that the differential entropy is closely related to the causal holographic information proposed in \cite{Hubeny1204} and studied in \cite{Kelly:2013aja}

In \cite{Myers}, the concept of Differential Entropy (\ref{RE}) has been modified and generalized to higher dimensions when the bulk curve has planar symmetry\footnote{Correspondingly, the boundary ``interval'' now becomes ``strip''.}. The reason for the modification  in higher dimension is that using the causal holographic information associated with the boundary strip leads to divergent results, though it works fine in AdS$_3$. The investigation in \cite{Myers} relied more on geometric construction. The basic idea is that bulk surface could be taken as the outer envelope of the bulk regions associated with the boundary intervals. It has been checked that the relation (\ref{equality}) holds  in various situations, including other backgrounds which is asymptotically $AdS$ and the Lovelock gravity.  Though the studies in \cite{Myers} strongly suggest that there is a new holographic equivalence between the gravitational entropy of bulk surface and the Differential entropy in the boundary, the discussion were made case by case. It would be interesting to see  why the equivalence (\ref{equality}) holds in these cases and when (\ref{equality}) could break down.

In this note, we present a brief proof of the equivalence (\ref{equality}) based on strong sub-additivity of the entanglement entropy and some general properties of quantum field theory.  The key ingredients in our discussion are two concepts, which are called as the Upper Differential Entropy (UDE) $E_{u}$
\be
E_u=\lim_{n\to\infty}\sum_{k=1}^n[S_{EE}(I_k)-S_{EE}(I_k\cap I_{k+1})]\label{URE}
\ee
 and the Lower Differential Entropy(LDE) $E_{l}$ respectively
 \be
 E_l=\lim_{n\to\infty}\sum_{k=1}^n[S_{EE}(I_k\cup I_{k+1})-S_{EE}(I_k)]\label{LRE}.
 \ee
Note that $E_u$ is just the Differential Entropy defined previously. Then we show that the three quantity $E_u, S_{gr},E_l$ satisfy the inequality
\be
E_u\ge S_{gr}\ge E_l\label{inequality}.
\ee
On the other hand, in quantum field theory, we can prove under quite reasonable  assumption that  UDE and LDE are actually equal
\be
E_u=E_l.\label{U=L}
\ee
If the holographic entanglement entropy is the same as the one in boundary CFT, as has been proved in the case of AdS$_3$/CFT$_2$ correspondence\cite{Faulkner:2013yia,Hartman:2013mia}, or at most differs a global factor in higher dimensional cases, the relation (\ref{U=L}) leads to the equivalence (\ref{equality}). This provides another point of view on the holographic equivalence (\ref{equality}).

The structure of the paper is as follows. In section 2 we show the inequality (\ref{inequality}) based on the strong subadditivity in the bulk. Then we  check the equality of (\ref{U=L}) in some simple cases in section 3. This makes us to be familiar to the concepts given in section 2.  After that, we prove the relation  (\ref{U=L}) from the quantum field theory side in section 4. Some discussion and conclusion will be presented in the last section.

\section{Gravitational entropy between bounds}
\begin{figure}
\centering
\includegraphics[height=9cm,width=12cm]{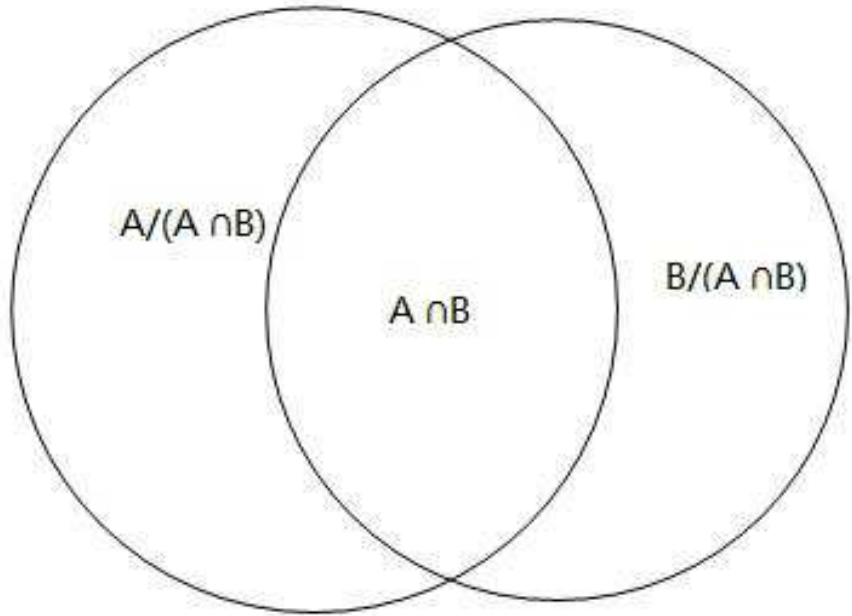}
\caption{An illustration of the strong subadditivity. The two circle are the regions $A$ and $B$, there can be some overlap between $A$ and $B$ and it is denoted as $A\cap B$.  }\label{ssb1}
\end{figure}

\begin{figure}
\centering
\includegraphics[height=3cm,width=12cm]{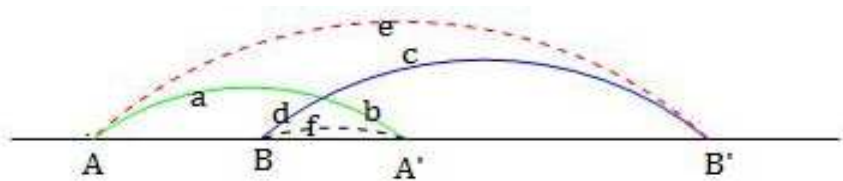}
\caption{Holographic proof of the strong subadditivity. The straight line $ABA'B'$ corresponds to the boundary space. The line $AA'$ is the region $A$ and $BB'$ is the region $B$. $AB'$ is the region $A\cup B$ and $BA'$ is the region $A\cap B$. According to the RT formula, the green curve,  blue curve, dashed red curve and dashed black curve represent the extremal surface which are homologous to $AA', BB',AB'$ and $BA'$ correspondingly. Their area are labeled as $a+b,c+d,e$ and $f$ correspondingly.  }\label{ssb2}
\end{figure}

In quantum information theory, there is a fundamental inequality \cite{Lieb73prl,Lieb73jmp} which is called strong subadditivity. One of its form is\footnote{Here and in the following, for simplicity we will use $S$ instead of $S_{EE}$ to denote the entanglement entropy.}
\be
S(A)+S(B)\ge S(A\cap B)+S(A\cup B)\label{ssb}
\ee
We  illustrate the inequality in Figure \ref{ssb1}. Since we are interested in the entanglement entropy in quantum field theory, the label $A$ and $B$ are chosen to be some spacelike region in the figure.  The original proof of strong subadditivity (\ref{ssb})  relies on some nontrivial properties of the entanglement entropy\cite{Lieb73jmp}. However, due to the RT formula, there indeed be a straightforward demonstration of this inequality\cite{Headrick0704}. The key point is shown in Figure \ref{ssb2}. It is easy to see
\be
S(A)+S(B)=(a+b)+(c+d)=(a+c)+(b+d)\ge e+f=S(A\cup B)+S(A\cap B)\label{proofssb}
\ee
The $\ge$ appear since $a+c$($b+d$) and $e$($f$) are homologous to  the same boundary $AB'$($BA'$) while $e$($f$) is the minimal surface according to RT formula. The holographic demonstration (\ref{proofssb}) is the simplest proof of the strong subadditivity we ever find. Here we use a similar argument to introduce the notions of UDE and LDE\footnote{The RT formula proposed in \cite{Ryu0603}mainly focus on static spacetime. The corresponding proof of strong subadditivity in \cite{Headrick0704} is only valid for static spacetime. A covariant formula is proposed in \cite{Hubeny:0705}, and the related work on strong subadditivity in this covariant framework can be found in \cite{Wall}. Our discussion in this paper focus on static spacetime.}.

\begin{figure}
\centering
\includegraphics[]{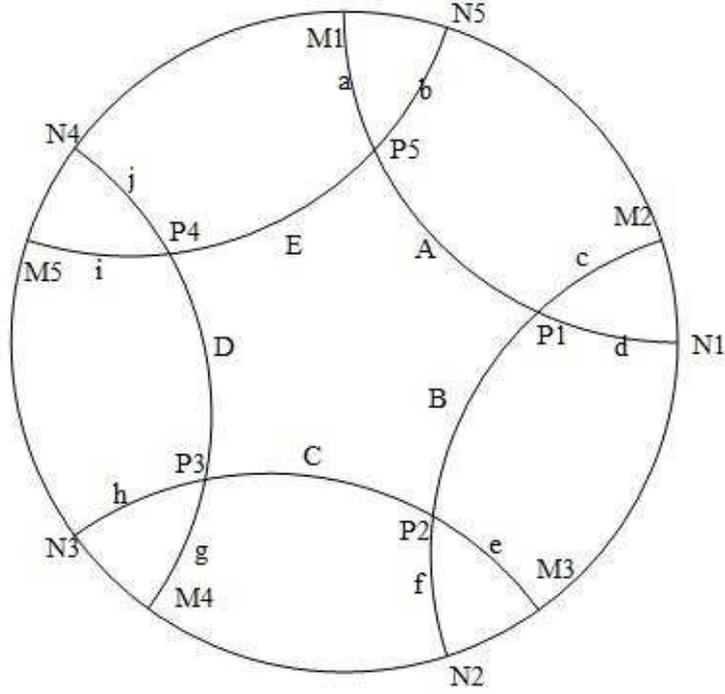}
\caption{An illustration of the appearance of UDE and LDE. We use five interval $I_i=M_iN_i(i=1,2,\cdots,5)$ to present the idea. We still use the symbol $M_iN_i$ to denote the corresponding bulk extremal surface without confusion. Each pair of consecutive bulk surface $M_iN_i, M_{i+1}N_{i+1}$ will meet at a codimension 3 surface which we denote as a point $P_i$ in the figure. Then there is an emergent bulk surface $P_1P_2\cdots P_5$ as shown in the figure. We have divide the extremal bulk surface as figure \ref{ssb2}. For example, the area of the bulk surface $M_1N_1$ has been divided to three terms a,A,d.   }\label{skeleton}
\end{figure}

As shown in Fig. \ref{skeleton}, we choose a series of intervals $I_i=M_iN_i$ and denote the entanglement entropy as $S_i$ correspondingly. According to the RT formula, the entropy is given by the area of a bulk minimal surface which has the same boundary as the interval. Hence we have\footnote{Here the coefficient $1/4G_N$ has been omitted to simplify the notation.}
\bea
&&S_1=a+A+d\nn\\
&&S_2=c+B+f\nn\\
&&S_3=e+C+h\\
&&S_4=g+D+j\nn\\
&&S_5=i+E+b\nn
\eea
In the boundary, we use $I_{1\cup2}$ to represent $I_1\cup I_2=M_1N_2$ and $I_{1\cap2}$ to represent $I_1\cap I_2=M_2N_1$. Then we find the inequality from the geometry
\bea
S_{1\cup2}\le a+A+B+f,\hs{3ex} S_{1\cap2}\le c+d\nn\\
S_{2\cup3}\le c+B+C+h,\hs{3ex} S_{2\cap3}\le e+f\nn\\
S_{3\cup4}\le e+C+D+j,\hs{3ex} S_{3\cap4}\le g+h\\
S_{4\cup5}\le g+D+E+b,\hs{3ex} S_{4\cap5}\le i+j\nn\\
S_{5\cup1}\le i+E+A+d,\hs{3ex} S_{5\cap1}\le a+b\nn
\eea
The bulk surface $P_1P_2P_3P_4P_5$ has an area $\tilde{S}$ as
\be
\tilde{S}=A+B+C+D+E
\ee
Then one easily find
\be
S_u\ge \tilde{S}\ge S_l
\ee
where $S_u$ and $S_l$ are defined to be
\be
S_u=\sum_{r=1}^{r=5} (S_r-S_{r\cap r+1}),\ S_l=\sum_{r=1}^{r=5}(S_{r\cup r+1}-S_r)
\ee
They are respectively called the Upper and Lower Differential Entropy of the surface $P_1P_2P_3P_4P_5$.
A general definition of the $n$-th Upper and the $n$-th Lower Differential Entropy are
\be
S_u^{(n)}=\sum_{r=1}^{r=n} (S_r-S_{r\cap r+1}),\ S_l^{(n)}=\sum_{r=1}^{r=n}(S_{r\cup r+1}-S_r)
\ee
The subscript $n$ means that there are $n$ intervals and $S_{n+1}=S_1$. In the limit $n\to \infty$, the bulk surface becomes smooth and the inequality should hold still,
\be
E_u\ge \tilde{S}_{\infty}\ge E_l.
\ee
We have used $E_u$ and $E_l$ to denote the limiting quantity of the $n$-th Upper and Lower Differential Entropy and they are the quantity UDE and LDE we have shown in the introduction
\be
E_u=\lim_{n\to\infty} S_u^{(n)},\ E_l=\lim_{n\to\infty} S_l^{(n)},
\ee
and $\tilde{S}_{\infty}$ is just the gravitational entropy $S_{gr}$
\be
\tilde{S}_{\infty}=S_{gr}
\ee
introduced  in the Introduction. Hence, we have proven the inequality (\ref{inequality}).

\section{Examples}

In the following examples, we  find that when the number of the intervals tends to infinity, the inequality actually becomes equality.
\be
E_u=S_{gr}=E_l.
\ee

\begin{figure}
\centering
\includegraphics[]{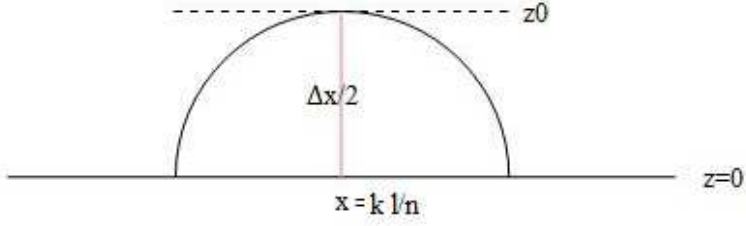}
\caption{$E_u=E_l$ when the bulk surface is $z=z_0$ }\label{ads3}
\end{figure}

The first example is the $AdS_3$. We choose the Poincar\'e coordinate and set the AdS radius to be unit
\be
ds^2=\frac{-dt^2+dx^2+dz^2}{z^2}.
\ee
 The boundary space is at infinity where $z=0$, with its length being $l\to\infty$.
When the bulk surface $S_{\infty}$ is a circle $z=z_0$, we use $n$ polygons to  approximate it. Hence, for every vertex $x_k=k l/n$, we can find an interval $I_k$ whose extremal surface in the bulk is tangent to the bulk surface at $(x_k,z_0)$. This is shown in Figure \ref{ads3}.

Since the extremal surface is a half circle $(x-x_k)^2+z^2=z_0^2(z>0)$, we find the length of the interval is
\be
\Delta x_k=2z_0.
\ee
For the interval $I_{k+1}$, its center is at $x_{k+1}=(k+1)l/n$ and the length is the same. Then the length of the intersection $I_k\cap I_{k+1}$ is
\be
\Delta x_{I_k\cap I_{k+1}}=2z_0-l/n.
\ee
The length of the union $I_k\cup I_{k+1}$ is
\be
\Delta x_{I_k\cup I_{k+1}}=2z_0+l/n.
\ee
Hence, we find the upper bound and the lower bound are respectively
\bea
S_u^{(n)}=n\frac{c}{3}\ln \frac{2z_0}{2z_0-l/n}\\
S_l^{(n)}=n\frac{c}{3}\ln \frac{2z_0+l/n}{2z_0}
\eea
where $c=\frac{3}{2G_N}$. In the limit $n\to\infty$, the upper and the lower bound approach to each other and equal to the area of the bulk surface $z=z_0$
\bea
E_u=E_l=S_{gr}=\frac{l}{4G_N z_0}.
\eea

\begin{figure}
\centering
\includegraphics{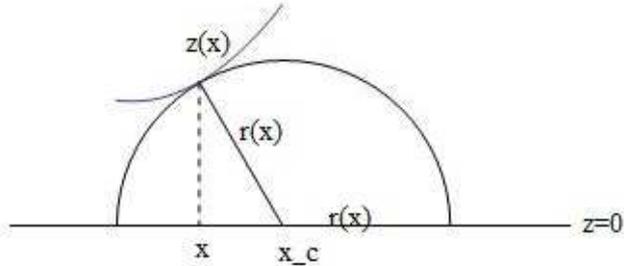}
\caption{$E_u=E_l$ when the bulk surface is $z=z(x)$}\label{ads3arb}
\end{figure}

In general, the bulk surface can be $z=z(x)$. In this case, we replace $k$'s with the coordinates $\cdots,x,x+dx\cdots$. Then for an interval $I_x$, we need to find its center point $x_c(x)$ and its length $\Delta x(x)=2r(x)$. This is shown in the Figure \ref{ads3arb}.
The natural candidate of the extremal curve should be tangent to the curve $z(x)$, hence we find
\be
x_c=x+z z'(x),\ r(x)=z\sqrt{1+z'^2}.
\ee
For the interval $I_{x+dx}$, we just replace $x$ to $x+dx$. Then
\bea
&&S(I_x)-\frac{1}{2}(S(I_x\cap I_{x+dx})+S(I_x\cap I_{x-dx}))\nn\\&\simeq& \frac{1}{2}(S(I_x\cup I_{x+dx})+S(I_x\cup I_{x-dx}))-S(I_x)\nn\\&\simeq&\frac{1}{4G_N}\frac{1+z'^2+zz''}{z\sqrt{1+z'^2}}dx+\mathcal{O}(dx^2)
\eea
and
\be
E_u=E_l=\frac{1}{4G_N}\int \frac{1+z'^2+zz''}{z\sqrt{1+z'^2}}dx=S_{\infty}(z=z(x))+\frac{1}{4G_N}\arcsin (z')|_0^l
\ee
We assume the bulk curve is smooth and periodic such that the boundary terms in the right hands side vanish.

 In the two examples given above, we find that the inequality from the strong subadditivity is actually an equality when the number of intervals tends to infinity. The fact that UDE and LDE are both equal to a bulk surface area holds for more general cases. One can check this point in all the examples given in \cite{Myers}.

From strong subadditivity, we only know that UDE is no smaller  than LDE. The fact that  the equality is actually saturated deserves interpretation. We will study this issue from the point of view of quantum field theory in the next section.

\section{$UDE=LDE$ in $QFT$}

In this section, we show that UDE and LDE are equal for a general local $QFT$\footnote{The $QFT$ we consider are continuous $QFT$. For lattice $QFT$, the equality found in this section is not valid.}. This relies on some simple assumptions which should hold for general $QFT$s. The entanglement entropy we consider corresponds to an interval in 2D QFT or an strip in $d>2$ QFT\footnote{We will use uniformly the word "interval"  even when it is a strip for higher dimensional quantum field theory.}. The intervals we place are in $x$ direction. The other spatial directions which are orthogonal to $x$ direction are written collectively as $y_i,i=2,3,\cdots,d-1$. All the spatial directions extend infinitely. However, we set an IR cutoff by restricting the length of each direction to be $l,l_2,l_3,\cdots,l_{d-1}$ respectively.  To construct the quantities UDE and LDE, we need an infinite set of intervals $I_x$. We use the point $x$ to label each interval as $I_x$, within it the center being $x_c(x)$ and the length being $\Delta x=2r(x)$. For consecutive intervals $I_x$ and $I_{x+dx}$, the center and the length of the interval should varies smoothly. When $x>l$, we require $I_x\sim I_{x+l}$.  In addition, we  assume that the $QFT$ satisfy two other requirements
\begin{enumerate}
\item $QFT$ is translational invariant.
\item As the intervals varies smoothly, we exclude the case that the entanglement entropy varies non-smoothly. In other words, the entanglement entropy (and all its derivatives\footnote{Actually, we only need it to be $C^2$-differentiable for our demonstration.}) is differentiable with respect to the interval.
\end{enumerate}
Since the $QFT$ is translational invariant, for an interval which is characterized by the boundary point $u$ and $v$, the entanglement entropy depends only  on the length of the interval $\Delta x=|u-v|$, namely $S_{EE}=S_{EE}(\Delta x)$. There are four kinds of intervals: $I_x$ and its neighbor $I_{x+dx}$, their intersection $I_x\cap I_{x+dx}$ and their union $I_x\cup I_{x+dx}$, which have the distances respectively
\bea
\Delta x(I_x)&=&2r(x),\\
\Delta x(I_{x+dx})&=&2r(x)+2r'dx+r''dx^2+\mathcal{O}(dx^3),\\
\Delta x(I_x\cap I_{x+dx})&=&2r(x)+(-x_c'+r')dx+\frac{1}{2}(r''-x_c'')dx^2+\mathcal{O}(dx^3),\\
\Delta x(I_x\cup I_{x+dx})&=&2r(x)+(x_c'+r')dx+\frac{1}{2}(r''+x_c'')dx^2+\mathcal{O}(dx^3).
\eea
 The prime in the superscript denotes the derivative with respect to $x$. Since $I_x\cap I_{x+dx}\subseteq I_x,\ I_x\cup I_{x+dx}\supseteq I_x$, we have
 \be
 r'^2-x_c'^2=(r'+x_c')(r'-x_c')\le0.
 \ee
As the entanglement entropy changes smoothly, we can safely do Taylor expansion for the entanglement entropy. We find that the mutual information
\bea
S(I_x,I_{x+dx})&=&S_x+S_{x+dx}-S_{x\cup x+dx}-S_{x\cap x+dx}\nn\\
&=&\frac{\p^2S(\Delta x)}{\p (\Delta x)^2}(r'^2-x_c'^2)dx^2+\mathcal{O}(dx^3)
\eea
vanishes as it is proportional to $(dx)^2$. Hence the upper and the lower bound should be equal. They are
\be
E_u=E_l=\int_0^l dx \frac{\partial S_{EE}}{\partial \Delta x}x_c(x)'.\label{URE}
\ee
Note that for the expansion should be consistent with the strong subadditivity $S(I_x,I_{x+dx})\ge 0$, we must have $\frac{\p^2S(\Delta x)}{\p (\Delta x)^2}\le 0$ as $r'^2-x_c'^2\le 0$.

 In the above discussion, we did not require the $QFT$ to be conformal invariant.  That means even though the theory is not a CFT, we can still find the equality $E_u=E_l$. For a theory which has a holographic dual, and the holographic entanglement entropy is given by the RT formula\footnote{Of course, we should demand the $QFT$ satisfy the previous two requirements. In the following, we always assume the two requirements to be satisfied when we consider a quantum field theory.},  since UDE and LDE are equal, there should be a surface whose area is the same as the two bounds when we take into account of the holographic inequality (\ref{inequality}) above. According to the RT formula, there is an emergent extremal surface corresponding to an entanglement entropy. Now, after one calculates the quantity $UDE$ (or LDE), there is a corresponding emergent bulk surface, which  need not to be extremal. Similar to the holographic entanglement entropy, the entropy of the surface could be understood as a generalized gravitational entropy as well.

\begin{figure}
\centering
\includegraphics{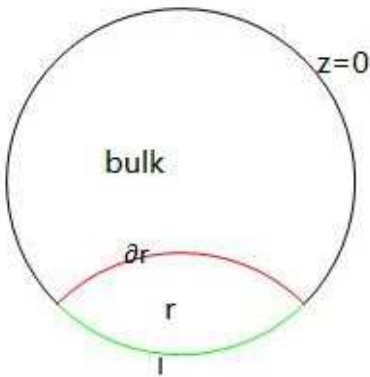}
\caption{The correspondence between boundary interval $I$ and bulk region $r$. The original RT formula provides an one to one correspondence between $I$ and $\partial r$. It is also an one to one correspondence between $I$ and $r$.}\label{extr}
\end{figure}

\begin{figure}
\centering
\includegraphics{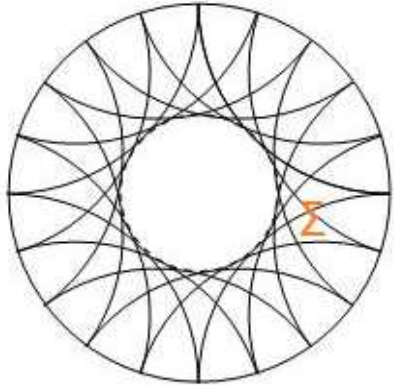}
\caption{ Bulk region $\Sigma$ from the boundary. $\Sigma$ is the region between the dashed circle and the boundary. }\label{circle case}
\end{figure}

 One may wonder why we need so many intervals  to reconstruct a bulk surface while for the extremal surface only one interval is needed. For an interval region $I$, we can always associate it a bulk region $r$, whose boundary in the bulk is $\partial r$, as we show in  Figure \ref{extr}.
After we specify the definite boundary $QFT$ and its bulk dual theory, the bulk region $r$ and the boundary region $I$ are in one to one correspondence\footnote{ There are similar statement in \cite{Headrick1312} and some earlier discussions on this issue  in \cite{Czech}.}. However, for a region $\Sigma$ bounded by a circle(and boundary), as we show in Figure \ref{circle case}, there is no single interval $I$ dual to it. However, the region $\Sigma$ can be constructed as the union of $r_k$ as shown in the figure,
   \be
    \mathop\bigcup_{k}^{\infty} r_k=\Sigma.\label{bulk}
   \ee\
   Each $r_k$ corresponds to an interval $I_k$ such that the region $\Sigma$ corresponds to an infinite number of intervals $\{\cdots,I_k,\cdots\}$. Then the gravitational entropy of the dashed circle, which is the bulk part of the boundary of $\Sigma$, must map to a boundary observable which is related to $\{\cdots,I_k,\cdots\}$. This observable in QFT has been shown to be the $UDE(LDE)$. For more general bulk region,  one can have
 \be
  \Sigma\subseteq\mathop\bigcup_{k}^{\infty} r_k.\label{bulk}
\ee
The most natural candidate of $I_k$ is constructed as we have discussed in the Introduction.

\begin{figure}
    \begin{center}
\includegraphics{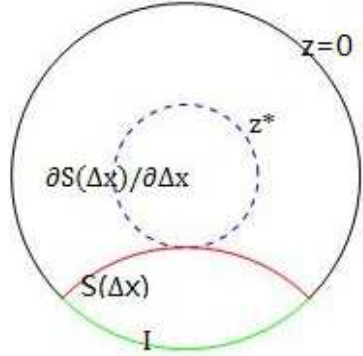}
\caption{An illustration of $\frac{\partial S_{EE}}{\partial \Delta x}$. $z=0$ is the boundary. $\Delta x$ is the length of the boundary interval which we use  green curve to represent it. The red curve is the bulk extremal surface corresponds to the interval. Its area is proportional to the entanglement entropy of the interval. The dashed blue circle $z=z*$ is determined by the interval in the boundary. The area of the bulk circle is proportional to the quantity $\frac{\partial S_{EE}}{\partial \Delta x}$. One should not be confused with the dimension matching problem, as is shown by (\ref{dEEformula}), $\frac{\partial S_{EE}}{\partial \Delta x}$ is actually the gravitational entropy density. }\label{pspx}
\end{center}
\end{figure}

 The formula (\ref{URE}) provides a new interpretation of the quantity $\frac{\partial S_{EE}}{\partial \Delta x}$. To simplify discussion, we choose $x_c=x$ and $\Delta x=const$, which corresponds to  the situation that the bulk surface is a circle. We immediately see that\footnote{Actually, (\ref{prop}) is only valid up to a totally derivative. However, we choose the bulk curve to be a circle, we expect the totally derivative vanishes.}
    \be
    \frac{\partial S_{EE}}{\partial \Delta x}\propto Area(z^*)\label{prop}
    \ee
    or more precisely,
    \be
    \frac{\partial S_{EE}}{\partial \Delta x}=\frac{1}{4G_N} \frac{Area(z=z^*)}{l}=\frac{S_{gr}(z=z^*)}{l}\label{dEEformula}
    \ee
    where $l$ in the denominator  is just the length in the $x$ direction. If we view $\frac{1}{4G_N} Area(z=z^*)$ as a generalized gravitational entropy, then $\frac{\partial S_{EE}}{\partial \Delta x}$ is the gravitational entropy density.
    This is illustrated by Figure \ref{pspx}.
 Roughly speaking, $\frac{\partial S_{EE}}{\partial \Delta x}$ has a geometrical meaning, as it is a measure of the area of the dual bulk circle. Several remarks are in order.
     \begin{enumerate}
     \item It is easy to see
     \be
     \frac{\partial S_{EE}}{\partial \Delta x}\ge0.\label{see>0}
     \ee
      This is obvious from the geometric meaning of $\frac{\partial S_{EE}}{\partial \Delta x}$, as the area is always non-negative. In the $QFT$ side, $\frac{\partial S_{EE}}{\partial \Delta x}$ is proportional to the central charge of the system, which counts the degrees of freedom of the system, so it should be non-negative.
      \item
      There are two possible ways to saturate the bound in (\ref{see>0}). The first one is to set the metric to be degenerate from the equation (\ref{dEEformula}). However, this is quite unusual so we will not consider this possibility. Another possible way
       to approach $0$ is when $\Delta x\to \infty$. In this limit, the dashed circle in Figure \ref{pspx} tends to a point if the bulk is global AdS. However, even though $\Delta x\to \infty$, the quantity $\frac{\partial S_{EE}}{\partial \Delta x}$ does not always tend to zero. When the background is an asymptotically $AdS$ black hole, for example, a $BTZ$ black hole,
       \be
       S_{gr}\to S_{BH},\ \Delta x\to \infty.\label{limit}
       \ee
       As $\Delta x\to \infty$, the extremal surface goes towards the horizon of the black hole. Combining (\ref{dEEformula}) with (\ref{limit}), one finds
    \be
    S_{EE}(\Delta x)\to s_{thermal}\Delta x,\ \Delta x\to \infty \label{EETh}
    \ee
    where $s_{thermal}$ is the thermal entropy density. This is consistent with the well known result.
    \item
     Let us give a precise realization of (\ref{dEEformula}). Choosing a general metric which is a holographic background,
    \be
    ds^2=-g_0(z)dt^2+g_1(z)dx^2+\sum_{i=2}^{d-1}g_i(z)dy_i^2+f(z)dz^2,
    \ee
    then we find
    \be
    \frac{\partial S_{EE}}{\partial \Delta x}=\frac{l_2\cdots l_{d-1}}{4G_N}G(z^*)
    \ee
    where $G(z)=\sqrt{g_1\cdots g_{d-1}}$. This is consistent with the result in \cite{Myers}.
    \item We can make use of the relation (\ref{dEEformula}) to reconstruct the black hole background. We take $AdS_3/CFT_2$ as the prototype. The one interval entanglement entropy in $CFT_2$ is  \cite{Holzhey,Calabrese}
        \be
        S(\Delta x)=\frac{c}{3}\ln \frac{\Delta x}{\delta}.\label{eeads3}
        \ee
        From holography, this comes from the minimal length of the curves which is homologous to the interval in the bulk of $AdS_3$ in Poincar\'e coordinate. After a conformal transformation, (\ref{eeads3}) becomes the finite temperature entanglement entropy
        \be
        S(\Delta x)=\frac{c}{3}\ln\frac{\beta}{\pi \delta}\sinh\frac{\pi \Delta x}{\beta}
        \ee
        As there is no rotation, we may assume the metric in the bulk to be\footnote{We do not have a good reason why the $g_{tt}$ should be the inverse(up to a minus sign) of $g_{rr}$ just from $CFT$ consideration yet. So we just assume it is like this.}
        \be
        ds^2=\frac{-dt^2}{f(r)}+f(r)dr^2+r^2 dx^2
        \ee
       The metric should be asymptotic $AdS_3$, so $f(r)\to \frac{1}{r^2}$ when $r\to\infty$. Then we can do power expansion of $f(r)$ around $r\to \infty$ as
      \be
       f(r)=\frac{1}{r^2}(1+\frac{\alpha_1}{r}+\frac{\alpha_2}{r^2}+\frac{\alpha_3}{r^3}+\frac{\alpha_4}{r^4}+\cdots).
           \ee
        From the RT formula, we find
         \be
          \Delta x=2\int_{r_0}^{\infty}\frac{dr r_0}{r^2\sqrt{r^2-r_0^2}}\sqrt{f(r)}\label{deltax1}
         \ee
where $r_0$ is the minimal value of the bulk coordinate where $r(x)'=0$\footnote{$r_0$ is similar to $z^*$ in Figure \ref{pspx}.}. On the other hand, from (\ref{dEEformula}),
\be
\frac{2\pi}{\beta}\frac{\cosh\frac{\pi\Delta x}{\beta}}{\sinh\frac{\pi\Delta x}{\beta}}=r_0,\label{deltax2}
\ee
we expand $\Delta x$ by the powers of $\frac{1}{r_0^n}$ in (\ref{deltax1},\ref{deltax2}), then the coefficient $\alpha_i$ are determined order by order
\be
\alpha_1=0,\ \alpha_2=\frac{4\pi^2}{\beta^2},\ \alpha_3=0,\ \alpha_4=\frac{16\pi^4}{\beta^4}\cdots
\ee
This determines $f(r)$ uniquely
\be
f(r)=\frac{1}{r^2-r_+^2},\ r_+=\frac{2\pi}{\beta}.
\ee
    \end{enumerate}

 The right hand side of (\ref{dEEformula}) is finite, hence for a general entanglement entropy of a strip
\be
S_{EE}=\frac{f_{d-2}}{\delta^{d-2}}+\cdots+\frac{f_{1}}{\delta}+p \ln \frac{g}{\delta}+ q, \label{EEstrip}
\ee
 the function $f_{d-2},\cdots,f_1,p$ should be independent of $\Delta x$. One can check this point for some simple $AdS$ cases\cite{Ryujhep}. In (\ref{EEstrip}) the last two terms are the universal terms, which are related to anomaly. The bulk surface which are only related to these universal terms deserves further study.

 If we define two new quantities
\bea
F_u=\lim_{n\to\infty}\sum_{k=1}^{n}\xi(I_{k})-\xi(I_k\cap I_{k+1})\\
F_l=\lim_{n\to\infty}\sum_{k=1}^{n}\xi(I_k\cup I_{k+1})-\xi(I_k)
\eea
where $\xi(I)$ is the causal information entropy, then we still find $F_u-F_l\to0$. However, as the causal information entropy does not satisfy the strong subadditivity in general\cite{Hubeny1204}, we cannot find a surface whose area is equal to them. This interprets why the causal information entropy is not a candidate in constructing the Differential Entropy\cite{Myers}.



\section{Discussion and Conclusion}

In this work, we have defined two bounds, Upper Differential Entropy $E_u$ and Lower Differential Entropy $E_l$,  in any quantum field theory. Both of them are constructed in terms of an infinite number of intervals(strips) which changes smoothly. When the quantum field theory is translational invariant and its entanglement entropy changes smoothly with respect to the length of the interval, the two bounds are actually equal. This has interesting implication when the quantum field theory has a gravitational dual. In the case that the RT formula is exact, we have shown that there is an emergent surface in the bulk whose gravitational entropy $S_{gr}$ falls exactly between the two bounds. In the infinite interval limit,  the three quantities must be equal. In this way, we have proven the holographic equivalence  found in
\cite{Balasubramanian1310} and extended in \cite{Myers}. The remarkable fact in the proof is that we just need some general properties of quantum field theory and the strong subadditivity of entanglement entropy. Another interesting corollary is encoded in the relation (\ref{dEEformula}), which intuitively interpret the quantity $\frac{\partial S_{EE}}{\partial \Delta x}$ to be the gravitational entropy density.

When the bulk theory is not the Einstein-Hilbert gravity, the area functional should change correspondingly. However, once the functional is extensive, the strong subadditivity is still satisfied\cite{Headrick0704}. In this case, one should replace $S_{gr}$ to a generalized area functional and then the identity (\ref{equality}) holds. However, there is indeed the case that the entropy  functional is not extensive. For example, the holographic higher spin entanglement entropy, which is proposed to be\cite{Boer, Ammon}
    \be
    S_{EE}(P,Q)=\frac{k_{cs}}{\sigma_{\frac{1}{2}}}\ln \lim_{\rho_0\to\infty} Tr_{\mathcal{R}}\mathcal{P}\exp(\int_Q^P\bar{A})\mathcal{P}\exp(\int_P^Q A)|_{\rho_Q=\rho_Q=\rho_0}
    \ee
     In the dual 2D CFT with W-symmetry, we can obtain the equality $E_u=E_l$ and their explicit forms from (\ref{URE}). However, it is not clear if  the functional is extensive for the bulk line which connects the point $P$ and $Q$. So whether the quantity $E_u$ or $E_l$  corresponds to some quantity in the bulk or not is a topic which deserves further exploration.

The proof of the equality $E_u=S_{gr}=E_l$ supports the idea that arbitrary bulk surface may connect to a gravitational entropy. However, this equality can only be proved when the boundary entanglement entropy is evaluated for the intervals or the strips. Correspondingly, the bulk surface $z=z(x)$  does not depend on other coordinates. For more general bulk surface, we have not found a suitable observable in the boundary theory. We believe some modification of the definition of $UDE$(or $LDE$) is inevitable as one cannot define two regions in succession in higher dimensions. For example, in three dimensional quantum field theory,  we need two coordinates $(x,y)$ to describe a general spacelike region. The interval $I_x$ in the definition of $UDE(LDE)$ need some generalization to $R_{x,y}$. There are at least two possible quantities $R_{x,y+dy}, R_{x+dx,y}$ analogue to $I_{x+dx}$. Consequently, we can construct out at least three entanglement entropies $S(R_{x,y}),S(R_{x+dx,y}),S(R_{x,y+dy})$. From the proof, $E_u=E_l$ is related to the fact that the mutual information $S(I_x,I_{x+dx})$ tends to zero in the second order of $dx$. So it seems that we need a more general definition of mutual information $S(R_{x,y},R_{x+dx,y},R_{x,y+dy})$. Since the mutual information is related to the strong subadditivity, maybe we need some generalization of strong subadditivity in order to include more objects. 

 As noted in \cite{Myers}, when $x_c'\le0$, the definition of Differential Entropy should be modified. Hence, in this case, we should exchange $\cap$ and $\cup$ in $UDE$ and $LDE$ for the interval $I_x$'s which obey $x_c'\le0$. The equality between $UDE$ and $LDE$ still holds.
\\\\

\noindent {\large{\bf Acknowledgments}} \\~\\
The work was in part supported by NSFC Grant No.~11275010, No.~11335012 and No.~11325522.

\end{document}